\documentclass[final,authoryear,3p,times,twocolumn]{vki_phd}

\usepackage{graphics}
\usepackage{graphicx}
\usepackage{epsfig}

\usepackage{amssymb}
\usepackage{amsmath}
\usepackage{pgf}
\usepackage{chemformula}
\usepackage{import}
\usepackage{cancel}
\DeclareUnicodeCharacter{2212}{-}
\DeclareUnicodeCharacter{2300}{$\varnothing$}
 \biboptions{square}

\journal{VKI}
\begin{document}

\begin{frontmatter}

%% Title, authors and addresses

%% use the tnoteref command within \title for footnotes;
%% use the tnotetext command for the associated footnote;
%% use the fnref command within \author or \address for footnotes;
%% use the fntext command for the associated footnote;
%% use the corref command within \author for corresponding author footnotes;
%% use the cortext command for the associated footnote;
%% use the ead command for the email address,
%% and the form \ead[url] for the home page:
%%
%% \title{Title\tnoteref{label1}}
%% \tnotetext[label1]{}
%% \author{Name\corref{cor1}\fnref{label2}}
%% \ead{email address}
%% \ead[url]{home page}
%% \fntext[label2]{}
%% \cortext[cor1]{}
%% \address{Address\fnref{label3}}
%% \fntext[label3]{}
%
%\title{}
%
%%% use optional labels to link authors explicitly to addresses:
%%% \author[label1,label2]{<author name>}
%%% \address[label1]{<address>}
%%% \address[label2]{<address>}
%
%\author{}
%
%\address{}

\title{Particle-based Simulation of an Air-Breathing Electric Propulsion System}
\author[label1,label2]{Pietro Parodi\corref{cor1}}
\author[label1]{Giovanni Lapenta}
\author[label2,label3]{Thierry Magin}
\address[label1]{Department of Mathematics, KU Leuven, Belgium}
\address[label2]{Aeronautics and Aerospace Department,  von Karman Institute for Fluid Dynamics, Belgium}
\address[label3]{Aero-Thermo-Mechanics Department, Universit\'e libre de Bruxelles, Belgium}
\cortext[cor1]{Corresponding author: \texttt{pietro.parodi@vki.ac.be}}

\begin{abstract}
A novel concept called Air-Breathing Electric Propulsion proposes to fly satellites at altitudes in the range 180-250~km, since this would have some advantages for the performance of radio communication and Earth observation equipment. The ABEP satellites compensate the atmospheric drag through a continuous thrust provided by collecting, ionizing and accelerating the residual atmospheric particles. It is clear that the feasibility of this concept will require a significant design and testing effort, performed first on ground and later in orbit. Plasma simulation tools play a fundamental role in the development of this technology, for two main reasons: (i) they can potentially increase dramatically the optimization and testing process of ABEP systems, since on-ground testing and in-orbit demonstrators are costly and time consuming, and (ii) the fidelity of on-ground testing is limited by the finite size and pumping speed of high-vacuum facilities, as well as the means through which the orbital flow is produced. In this paper, we demonstrate a one-way coupled, particle-based simulation strategy for a CubeSat sized ABEP system. The neutral flow in the full geometry of the ABEP system comprising the intake and the thruster is simulated first through Direct Simulation Monte Carlo. Then, the resulting neutral density is used as the input for a Particle-in-Cell simulation of the detailed thruster geometry. The simulations are performed in 3D and within the VKI in-house code \textsc{Pantera}, taking advantage of the fully-implicit energy-conserving scheme.
\end{abstract}

\begin{keyword}
Particle-in-Cell, Direct Simulation Monte Carlo, particle methods, implicit, energy-conserving, collisions, ion thruster.
\end{keyword}

\end{frontmatter}

% \linenumbers

%% main text

\section{Introduction}

Air-Breathing Electric Propulsion (ABEP) proposes to use atmospheric particles as the propellant for spacecraft in Very Low Earth Orbit instead of propellant stored on-board at launch. In such configuration, the plasma thruster is interfaced with a collector-intake. This is a device that collects particles from the ram-facing side of the satellite and takes them to the thruster ionization chamber, while increasing the density through passive or active means. Presently, no fully integrated ABEP system has ever been tested in orbit on in a truly representative environment, and the optimization process has still to be fully understood due to the complex interaction between subsystems \cite{zheng2020,andreussi2022}.

Designing and testing traditional EP systems is already a challenging and time consuming task. Vacuum chambers are fairly expensive to build and to operate, and any modification of the thruster or in the testing setup requires opening the vacuum chamber and pumping it down again. Additionally, the ground testing vacuum environment is not completely faithful to the orbital conditions, for various reasons. Even the largest facilities with very high pumping speed capability cannot reach the low background pressure present in space. The environment in terms of electric fields is not fully reproduced either, because of the presence of the metallic walls of the vacuum facility. Finally, in the case of ABEP, it is not currently possible to accurately reproduce experimentally the hypersonic, rarefied orbital flow encountered in orbit. These issues cause a discrepancy between the performance measured on ground and in orbit. For these reasons, simulation represents an essential tool for the further development of Electric Propulsion and ABEP technology \cite{tr-avt-294}. First, it can be used to obtain a deeper understanding of the driving physics and processes in a complex system such as ABEP. Then, it plays an important role in the design and optimization phase of such systems. Finally, it is necessary to correctly interpret and characterize on-ground experiments, which remain a fundamental step in the testing and qualification process, but also to predict the in-orbit performance by removing the effects and limitations of ground testing facilities.

The challenges in the numerical simulation of the type of flow involved in an ABEP system mainly originate from the large disparity of length and time scales involved. First, the collisional length scale, characterized by the mean free path $\lambda$, is typically larger or comparable to the size of the geometry. This results in a state of nonequilibrium of the translational degrees of freedom of the particles. For this reason we use particle-based methods, which can accurately and efficiently represent nonequilibrium velocity distributions. The large mass and temperature disparity between ions and electrons also causes these particles to move and oscillate on time scales that can differ by more than three orders of magnitude, all of which typically need to be resolved by the simulation. Methods have been independently developed to treat rarefied gases and plasmas. The Direct Simulation Monte Carlo (DSMC) method \cite{bird1963,bird1994} models particle-particle collisions such that the gas reproduces accurately macroscopic transport properties such as viscosity and diffusivity as it approaches thermal equilibrium. The Particle\nobreakdash-in\nobreakdash-Cell (PIC) method \cite{hockney} models the interaction of charged species through the electromagnetic fields they produce, as well as those that are externally generated. Typically, when in a PIC simulation collisions with a higher density background gas need to be taken into account, this is done through a method called Monte Carlo Collisions (MCC). The DSMC and PIC(-MCC) methods are compatible and can easily be merged in a hybrid PIC-DSMC algorithm. We implemented these methods in the \textsc{Pantera} software, developed at VKI.

In the present work, we demonstrate a strategy to simulate a 3U CubeSat-sized ($33\times 10 \times 10$~cm$^{3}$) ABEP system, where the target geometry is accurately reproduced in 3D, and the flow is accurately modeled starting from models that are close to first principles. We wish to demonstrate that particle-based simulation is a viable tool to produce predictions of engineering interest, and can be integrated into the design and testing process, with a role similar to that played by modern Computational Fluid Dynamics (CFD) in an industrial setting.

In Section~\ref{sec:methodology}, we will introduce the one-way coupling strategy between neutral flow in the ABEP system and plasma flow in the thruster. We will also detail the implementation of some relevant models and algorithms implemented in \textsc{Pantera}. In Section\ref{sec:domain}, we describe the simulation domain and the boundary conditions applied. Results are presented in Section~\ref{sec:results}, first for the neutral flow and then for the plasma flow in the ionization chamber and thruster plume. As a sanity check, a comparison with a global mass and power balance is performed in Section~\ref{sec:globalmodel}.

\section{Methodology}
\label{sec:methodology}

In this work, we follow a one-way coupled approach for the simulation of the ABEP system. The flow of the neutral atmospheric particles is simulated first, considering the full geometry of the intake and the thruster. Then, the result is used as a boundary condition to simulate the ionization and acceleration of the plasma in the thruster. In the following subsections we will detail the procedures for the two simulations.

\subsection{Neutral flow: DSMC}

The density of the residual atmosphere is such that, once the gas is compressed in the intake, it is in the so-called transitional flow regime, in which the mean free path $\lambda$ is comparable to the size of the intake. We simulate this flow using the DSMC method. In this method, particle motion and collisions are decoupled in time. Particles move in rectilinear trajectories, interacting with solid walls when one is encountered. Then, in each cell of the mesh a certain number of particle pairs is selected to collide. Collisions instantaneously change the velocity vector of the particles. After computing the desired moments of the particle distribution, the procedure is repeated for the next time step. The details of the algorithm can be found in \cite{bird1994,boydandschwartzentruber}. In \textsc{Pantera}, we use an unstructured simplex mesh, composed of triangles in 2D and of tetrahedra in 3D. This is very advantageous when devices of engineering interest have to be simulated, since it greatly simplifies mesh generation and adaptation for complex geometries.

\subsection{Plasma flow: PIC-MCC}

The ion thruster operates by ionizing neutral particles and accelerating them across a set of electrically biased grids. Electrons gain the energy necessary for ionization through the electric field induced by the coil that surrounds the ionization chamber. The simulation method must be able to accurately represent the resulting nonequilibrium, collisional plasma. In this work, we employ the Particle-in-Cell method, including collisions with a background gas through a method called Monte Carlo Collisions. Poisson's equation is discretized using the Finite Element Method on a first order basis built on the unstructured mesh. Notice that the mesh used to solve the field is also used to sort the particles and compute collisions. Here, we use the fully-implicit algorithm described in \cite{parodiphdsymp2023}, with a simplified Jacobian that is analogous to the semi-implicit treatment of \cite{lapenta2017}.

\subsubsection{Crank-Nicholson particle mover with a magnetic field}

In this work, the particle mover already presented in \cite{parodiphdsymp2023} is modified to consider the magnetic field generated by the current in the RF coil. The particles equations of motion (Newton-Lorentz), discretized using a Crank-Nicholson temporal scheme, read:

\begin{equation} \label{eq:moverposition}
    \mathbf{x}^{\nu+1} = \mathbf{x}^\nu + \mathbf{v}^{\nu+\frac{1}{2}} \Delta t,
\end{equation}
\begin{equation}
    \mathbf{v}^{\nu+1} = \mathbf{v}^\nu + \frac{q}{m}(\mathbf{E} + \mathbf{v}^{\nu+\frac{1}{2}} \times \mathbf{B}) \Delta t,
\end{equation}
where we called $\alpha = \frac{1}{2}\frac{q}{m}\Delta t$ and $\mathbf{a}=\mathbf{v}^\nu + \alpha \mathbf{E}$. Since $\mathbf{v}^{\nu+\frac{1}{2}} = (\mathbf{v}^\nu + \mathbf{v}^{\nu+1})/2$, we can write:
\begin{equation}
    \mathbf{v}^{\nu+\frac{1}{2}} = \mathbf{a} + \mathbf{v}^{\nu+\frac{1}{2}} \times \alpha \mathbf{B}
\end{equation}
Chen and Chac\'on have shown \cite{chen2023} that this equation can be solved analytically, resulting in:
\begin{equation}
    \mathbf{v}^{\nu+\frac{1}{2}} = \frac{\mathbf{a} + \mathbf{a} \times \alpha \mathbf{B} + (\mathbf{a} \cdot \alpha \mathbf{B}) \alpha \mathbf{B} }{1+(\alpha |\mathbf{B}|)^2}
\end{equation}
In order to maintain the accuracy of the parabolic trajectory integrator presented in \cite{parodiphdsymp2023}, we re-write the position update, Eq.~(\ref{eq:moverposition}) gathering the linear and quadratic terms in $\Delta t$:
\begin{multline} \label{eq:magnetizedmover}
    \mathbf{x}^{\nu+1} = \mathbf{x}^\nu + \Delta t \underbrace{ \left[ (1-\mathcal{B})  \mathbf{v}^\nu  + \mathcal{B} \left( \frac{ \mathbf{E}\times \mathbf{B} }{ |\mathbf{B}|^2 } + (\mathbf{v}^\nu \cdot \mathbf{b}) \mathbf{b} \right) \right] }_{\mathbf{v_\text{mover}(\alpha |\mathbf{B}|)}} \\ + \frac{1}{2}\Delta t^2 \underbrace{ \left[ (1-\mathcal{B}) \frac{q}{m}(\mathbf{E} + \mathbf{v}^\nu \times \mathbf{B}) + \mathcal{B} \frac{q}{m}(\mathbf{E} \cdot \mathbf{b}) \mathbf{b} \right]}_{\mathbf{a_\text{mover}(\alpha |\mathbf{B}|)}},
\end{multline}
where $\mathcal{B} = (\alpha |\mathbf{B}|)^2 /[ 1 + (\alpha |\mathbf{B}|)^2 ]$ is a parameter describing the level of magnetization of the particle, since $\alpha |\mathbf{B}| = \omega_c \Delta t/2$, with $\omega_c$ the cyclotron frequency. The parameter $\mathcal{B}$ tends to 1 in the fully magnetized limit, and to 0 in the unmagnetized limit. In principle, also $\mathcal{B}$ depends on the time of motion of the particle, meaning that Eq.~(\ref{eq:magnetizedmover}) is not simply quadratic in $\Delta t$, and finding the final particle position would actually require an iterative solution. In order to simplify the procedure, we compute $\mathcal{B}$ for all particles assuming they are moved for the full time step. Notice that in the fully magnetized limit, particles move and accelerate only along magnetic field lines while drifting according to the $\mathbf{E} \times \mathbf{B}$ drift. As noted by the authors in \cite{chen2023}, this mover does not include a $\nabla \mathbf{B}$-drift term in the asymptotic limit.

\subsubsection{Applied electromagnetic fields from the RF coil} \label{sec:rffield}

We simulate an RF coil wrapped around the ionization chamber driven with an alternating current $\tilde{I}_\text{coil} = I_\text{coil}\cos(2\pi f \, t)$ at a frequency $f=13.56$~MHz. We assume that the plasma does not modify the electric and magnetic field produced by the coil. This is referred to as the low-density limit. According to \cite{chabertbook}, in this case the field produced by the coil consists in an uniform axial magnetic field:
\begin{equation}
    \tilde{B}_x = \tilde{B}_{x\,0} = \frac{N}{\ell} I_\text{coil} \sin(2\pi f \, t)
\end{equation}
and an azimuthal electric field, $\pi/2$ out of phase, proportional to the radial distance $r$ from the axis:
\begin{equation}
    \tilde{E}_\theta = -\mu_0 \pi f \frac{N}{\ell} \cos(2\pi f \, t) \, r
\end{equation}
These fields are applied only to particles in the ionization chamber volume. The electric field is superimposed to the electrostatic field resulting from the charge distribution, as described by Maxwell's equations:
\begin{equation}
    \mathbf{E} = -\nabla \phi + \tilde{\mathbf{E}}.
\end{equation}
In fact, taking the divergence, we recover Poisson's equation:
\begin{equation}
    \nabla \cdot \mathbf{E} = -\nabla^2 \phi + \cancel{\nabla \cdot \tilde{\mathbf{E}}} = \frac{\rho}{\varepsilon_0},
\end{equation}
since the induced electric field $\tilde{\mathbf{E}}$ also satisfies Gauss's law and is therefore divergence-free. We compute the power transferred to the charged particles as:
\begin{equation}
    \mathcal{P}_\text{abs} = \sum_p q_p W_p v_{p \, \theta} \tilde{E}_\theta(\mathbf{x}_p).
\end{equation}
We've found that simply fixing a value for $I_\text{coil}$ causes the plasma to either extinguish or to grow exponentially in density. To remedy this, inspired by the procedure adopted in \cite{maginthesis}, in the simulation the coil current is adjusted once every RF cycle, based on the target absorbed power $\mathcal{P}_\text{tgt}$ and the average over the previous RF cycle of the absorbed power, $\mathcal{P}_\text{avg}$, using the following formula:
\begin{equation} \label{eq:icoil}
    I_\text{coil} = I_\text{coil, old} \sqrt{\frac{\mathcal{P}_\text{tgt}}{\mathcal{P}_\text{avg}}}
\end{equation}
We've found this to be sufficient to stabilize the set value of the current.

\subsubsection{MCC Collisions with the neutral background}

When considering the plasma in the thruster, we assume that ion-neutral and electron-neutral collisions occur much more frequently than collisions between ions and electrons. This is a common assumption in plasmas with a low ionization degree. Then, collisions are only computed between simulated particles and a fixed background of neutrals. We implemented the algorithm of Vahedi \cite{vahedi1995}, with a slight modification. As in the original algorithm, a ``null" collision probability is computed as follows:
\begin{equation}
    P_\text{null} = 1-\exp(-\nu' \Delta t)
\end{equation}
Each particle is selected to collide with this probability, then the probability of it undergoing one of the possible processes (elastic collisions, chemical reactions, or a null collision)  is selected using:
\begin{align*}
    & R \le \tfrac{P_1(\mathcal{E})}{P_\text{null}} \quad &\text{(proc. 1)}\\
    \tfrac{P_1(\mathcal{E}_i)}{P_\text{null}} <\,  & R \le \tfrac{P_1(\mathcal{E}) + P_2(\mathcal{E})}{P_\text{null}} \quad &\text{(proc. 2)}\\
    & \vdots \\
    \textstyle \tfrac{\sum_{j=1}^N P_j(\mathcal{E})}{P_\text{null}} <\, & R \quad &\text{(null)}
\end{align*}
Where the probability of each process depends on the corresponding cross-section, which is function of the available kinetic energy $\mathcal{E}$ in the collision reference frame:
\begin{equation}
    P_j(\mathcal{E}) = 1- \exp[-n_\text{BG} \sigma_j(\mathcal{E}) g \Delta t],
\end{equation}
where $n_\text{BG}$ is the background number density, $\sigma_j$ is the cross section for process $j$, and $g$ is the relative velocity between the colliding particles. This slightly more complex form with respect to what originally proposed by Vahedi allows to remove the dependence of the number of particles undergoing a (non-null) collision process on the value chosen for $\nu'$. The latter then becomes merely a parameter to optimize the efficiency of the algorithm, whose particular value does not affect the result.

\begin{figure}
    \centering
    \input{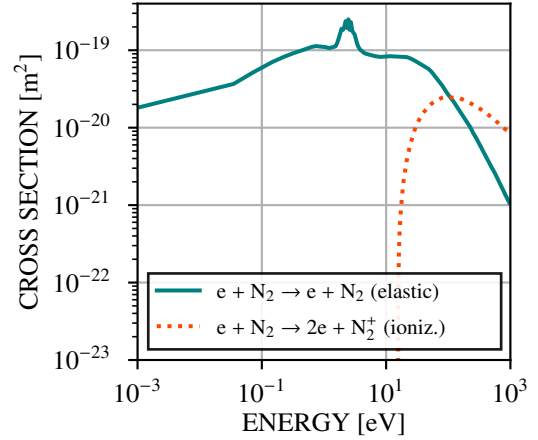}
    \caption{Elastic and ionization cross-sections for electron impact with nitrogen molecules. Tabulated data from the LXCat database \cite{lxcatbiagi}.}
    \label{fig:crossections}
\end{figure}

We have selected as collisional processes the elastic scattering of electrons and electron impact ionization. The reaction set is a subset of what is adopted by Taccogna et al. in \cite{taccogna2022}. These are the main drivers of electron energy loss and plasma generation. The cross sections, plotted in Figure \ref{fig:crossections}, are extracted from the Biagi dataset, retrieved from LXCat \cite{lxcatbiagi}. The activation energy for ionization is 15.581~eV. A more complete reaction set for a mixture of nitrogen and oxygen is foreseen in the future, including dissociation and excitation reactions, which represent additional energy loss mechanisms.

\subsection{Simulation domain, mesh, and boundary conditions} \label{sec:domain}

\begin{figure}[ht]
    \centering
    \def\svgwidth{\columnwidth}
    \begin{tiny}
    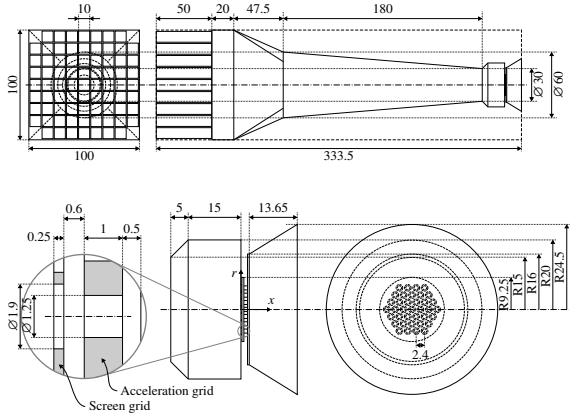
    \end{tiny}
    \caption{Dimensions in [mm] of the full Air-Breathing system geometry (top) and of the thruster geometry (bottom), with a detail of the grid apertures (bottom left). The full geometry is sized for a 3U CubeSat, taking inspiration from \cite{gaffarelstp}, and the thruster is inspired by the RIT-\textmu X thruster of ArianeGroup \cite{binder2017, binder2019}.}
    \label{fig:dimensions1}
\end{figure}

The simulation domain for the simulation of the neutral gas corresponds to the coupled ABEP intake and thruster, sized for a 3U CubeSat (10 $\times$ 10 $\times$ 33~cm). The design of the intake has been inspired to the work of J. Gaffarel \cite{gaffarelstp}. The dimensions are shown in Figure~\ref{fig:dimensions1}.

A mixture of 1~\ch{N2}+1~\ch{O} is injected at the inlet of the ducts with total number density $n_\infty = 4.2\times 10^{17}$~m$^{-3}$, velocity $v_\infty = 7800$~m/s and translational temperature $T_\infty=700$~K. Particle scattering at the walls is not fully accommodated in the tangential direction, following the Cercignani-Lampis-Lord (CLL) model, \cite{lord1991}, with accommodation coefficients $\sigma_n = 1.0$, $\sigma_t = 0.9$. Oxygen atoms recombine at the surface into molecular oxygen with perfect cataliticy, while particles reaching the boundary of the domain simply vanish. Collisions between particles are simulated with the Variable-Soft Sphere model, as described in Section~\ref{sec:domain}. The coefficients for the model are given in Table~\ref{tab:vsscoefficients}.

\begin{table}[htb]
\begin{center}
\caption{Species-specific parameters adopted for the Variable-Soft Sphere collision model in the DSMC simulation. The relative velocity-dependent collision diameter is computed from the reference values $d_\text{ref}$, $T_\text{ref}$ and the exponential factor $\omega$. Parameter $\alpha$ influences the anisotropicity of the post-collision scattering. The parameters are averaged for particle pairs of unequal species.} \label{tab:vsscoefficients}
\vspace{2mm}
\begin{tabular}{lcccc}
\hline
\hline
Species & $d_\text{ref}$        & $\omega$ & $T_\text{ref}$ & $\alpha$ \\
\hline
O$_2$   & $3.96\times 10^{-10}$ & 0.77     & 273.15         & 1.4      \\
N$_2$   & $4.07\times 10^{-10}$ & 0.74     & 273.15         & 1.6      \\
O       & $3.00\times 10^{-10}$ & 0.8      & 273.15         & 1        \\
\hline
\hline
\end{tabular}
\end{center}
\end{table}

\begin{figure}[ht]
    \centering
    %% Creator: Inkscape inkscape 0.92.3, www.inkscape.org
%% PDF/EPS/PS + LaTeX output extension by Johan Engelen, 2010
%% Accompanies image file 'thruster_boundcond.pdf' (pdf, eps, ps)
%%
%% To include the image in your LaTeX document, write
%%   \input{<filename>.pdf_tex}
%%  instead of
%%   \includegraphics{<filename>.pdf}
%% To scale the image, write
%%   \def\svgwidth{<desired width>}
%%   \input{<filename>.pdf_tex}
%%  instead of
%%   \includegraphics[width=<desired width>]{<filename>.pdf}
%%
%% Images with a different path to the parent latex file can
%% be accessed with the `import' package (which may need to be
%% installed) using
%%   \usepackage{import}
%% in the preamble, and then including the image with
%%   \import{<path to file>}{<filename>.pdf_tex}
%% Alternatively, one can specify
%%   \graphicspath{{<path to file>/}}
%% 
%% For more information, please see info/svg-inkscape on CTAN:
%%   http://tug.ctan.org/tex-archive/info/svg-inkscape
%%
\begingroup%
  \makeatletter%
  \providecommand\color[2][]{%
    \errmessage{(Inkscape) Color is used for the text in Inkscape, but the package 'color.sty' is not loaded}%
    \renewcommand\color[2][]{}%
  }%
  \providecommand\transparent[1]{%
    \errmessage{(Inkscape) Transparency is used (non-zero) for the text in Inkscape, but the package 'transparent.sty' is not loaded}%
    \renewcommand\transparent[1]{}%
  }%
  \providecommand\rotatebox[2]{#2}%
  \newcommand*\fsize{\dimexpr\f@size pt\relax}%
  \newcommand*\lineheight[1]{\fontsize{\fsize}{#1\fsize}\selectfont}%
  \ifx\svgwidth\undefined%
    \setlength{\unitlength}{134.39157486bp}%
    \ifx\svgscale\undefined%
      \relax%
    \else%
      \setlength{\unitlength}{\unitlength * \real{\svgscale}}%
    \fi%
  \else%
    \setlength{\unitlength}{\svgwidth}%
  \fi%
  \global\let\svgwidth\undefined%
  \global\let\svgscale\undefined%
  \makeatother%
  \begin{picture}(1,1.2095037)%
    \lineheight{1}%
    \setlength\tabcolsep{0pt}%
    \put(0,0){\includegraphics[width=\unitlength,page=1]{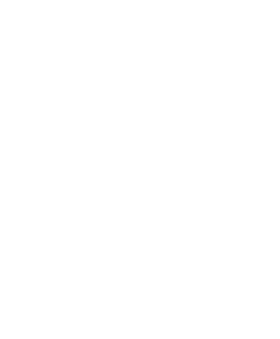}}%
    \put(0.45205149,0.09904852){\color[rgb]{0,0,0}\makebox(0,0)[lt]{\lineheight{1.25}\smash{\begin{tabular}[t]{l}$I_\text{coil}$\end{tabular}}}}%
    \put(0,0){\includegraphics[width=\unitlength,page=2]{thruster_boundcond.pdf}}%
    \put(0.08978113,0.27618479){\color[rgb]{0,0,0}\makebox(0,0)[rt]{\lineheight{1.25}\smash{\begin{tabular}[t]{r}$V_\text{RF}$\end{tabular}}}}%
    \put(0,0){\includegraphics[width=\unitlength,page=3]{thruster_boundcond.pdf}}%
    \put(0.53841884,0.96672928){\color[rgb]{1,0.16470588,0.16470588}\makebox(0,0)[lt]{\lineheight{1.25}\smash{\begin{tabular}[t]{l}$B_x$\end{tabular}}}}%
    \put(0.44400964,0.74816413){\color[rgb]{0,0.4,1}\makebox(0,0)[lt]{\lineheight{1.25}\smash{\begin{tabular}[t]{l}$E_\theta$\end{tabular}}}}%
    \put(0.37584175,1.00714019){\color[rgb]{0,0,0}\makebox(0,0)[rt]{\lineheight{1.25}\smash{\begin{tabular}[t]{r}$r$\end{tabular}}}}%
    \put(0.37494465,0.8407247){\color[rgb]{0,0,0}\makebox(0,0)[rt]{\lineheight{1.25}\smash{\begin{tabular}[t]{r}$\theta$\end{tabular}}}}%
  \end{picture}%
\endgroup%

    \caption{Boundary conditions for the field applied to the thruster. We model the RF field generated by a coil that surrounds the ionization chamber. The field is applied only in the shaded area. It is constituted of an azimuthal electric field linearly increasing with radius, and a uniform axial magnetic field. The screen grid and internal thruster walls are fed with an imposed alternating voltage, while the acceleration grid and external surfaces are grounded.}
    \label{fig:thruster_boundcond}
\end{figure}

The simulation domain for the plasma is restricted to the thruster, but the mesh has a much higher degree of refinement, in order to represent more accurately the fields in proximity of the more minute geometrical features such as the apertures of the grids, and to reasonably resolve the plasma sheaths on boundary surfaces. Initially, the ionization chamber is filled with a uniform, quasineutral plasma at $1\times 10^{17}$~m$^{-3}$. We consider a thruster where the grids are biased with an alternating voltage. This allows an alternate emission of ions and electrons, and therefore cathode-less operation, as described in \cite{rafalskyi_patent} and demonstrated in \cite{rafalskyi2014}. Further numerical investigation can be found in \cite{lafleur2019}. Notice that the reliability of such system has been put into doubt \cite{fu2021} due to the RF biased grid operation. However, the original proponents have already solved some of the technical limitations and demonstrated successful operation of the device on ground. The acceleration grid is grounded, as well as all other external surfaces of the thruster and the outlet plane. A voltage $\tilde{V}_\text{RF}=V_\text{DC} + \frac{1}{2} V_\text{pp} \, \sin(2\pi f \, t)$ is applied to the screen grid and the internal surfaces of the thruster. While this grid would normally be decoupled through a capacitor and allowed to self-bias, in our case the DC voltage is imposed. In the simulation, $V_\text{pp} = 1500$~V, $V_\text{DC} = 730$~V, and $f=13.56$~MHz, as for the RF coil. A schematic is shown in Figure~\ref{fig:thruster_boundcond}, which also shows the applied electric and magnetic fields from the coil described in Section~\ref{sec:rffield}.

\section{Results and discussion} \label{sec:results}

Here we present first the results for the simulation of neutral particles in Section~\ref{sec:neutralflow}, and then for the ionized flow, focusing on the phenomena happening in the ionization chamber in Section~\ref{sec:ionizchamber} and in the grid and external plume in Section~\ref{sec:thrusterplume}. Finally, in Section~\ref{sec:globalmodel}, we compare the plasma properties in the ionization chamber to those obtained from a global energy and mass balance calculation.

\subsection{Neutral flow} \label{sec:neutralflow}

\begin{figure}[ht]
    \centering
    %% Creator: Inkscape inkscape 0.92.3, www.inkscape.org
%% PDF/EPS/PS + LaTeX output extension by Johan Engelen, 2010
%% Accompanies image file 'neutrals_1.pdf' (pdf, eps, ps)
%%
%% To include the image in your LaTeX document, write
%%   \input{<filename>.pdf_tex}
%%  instead of
%%   \includegraphics{<filename>.pdf}
%% To scale the image, write
%%   \def\svgwidth{<desired width>}
%%   \input{<filename>.pdf_tex}
%%  instead of
%%   \includegraphics[width=<desired width>]{<filename>.pdf}
%%
%% Images with a different path to the parent latex file can
%% be accessed with the `import' package (which may need to be
%% installed) using
%%   \usepackage{import}
%% in the preamble, and then including the image with
%%   \import{<path to file>}{<filename>.pdf_tex}
%% Alternatively, one can specify
%%   \graphicspath{{<path to file>/}}
%% 
%% For more information, please see info/svg-inkscape on CTAN:
%%   http://tug.ctan.org/tex-archive/info/svg-inkscape
%%
\begingroup%
  \makeatletter%
  \providecommand\color[2][]{%
    \errmessage{(Inkscape) Color is used for the text in Inkscape, but the package 'color.sty' is not loaded}%
    \renewcommand\color[2][]{}%
  }%
  \providecommand\transparent[1]{%
    \errmessage{(Inkscape) Transparency is used (non-zero) for the text in Inkscape, but the package 'transparent.sty' is not loaded}%
    \renewcommand\transparent[1]{}%
  }%
  \providecommand\rotatebox[2]{#2}%
  \newcommand*\fsize{\dimexpr\f@size pt\relax}%
  \newcommand*\lineheight[1]{\fontsize{\fsize}{#1\fsize}\selectfont}%
  \ifx\svgwidth\undefined%
    \setlength{\unitlength}{167.24805216bp}%
    \ifx\svgscale\undefined%
      \relax%
    \else%
      \setlength{\unitlength}{\unitlength * \real{\svgscale}}%
    \fi%
  \else%
    \setlength{\unitlength}{\svgwidth}%
  \fi%
  \global\let\svgwidth\undefined%
  \global\let\svgscale\undefined%
  \makeatother%
  \begin{picture}(1,1.46850382)%
    \lineheight{1}%
    \setlength\tabcolsep{0pt}%
    \put(0,0){\includegraphics[width=\unitlength,page=1]{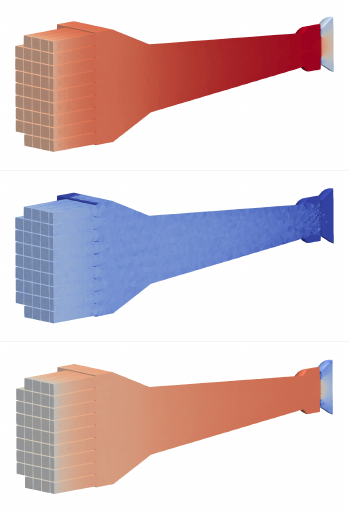}}%
    \put(0.71653449,1.1156759){\color[rgb]{0,0,0}\makebox(0,0)[lt]{\lineheight{1.25}\smash{\begin{tabular}[t]{l}N$_2$\end{tabular}}}}%
    \put(0.71653449,0.6223962){\color[rgb]{0,0,0}\makebox(0,0)[lt]{\lineheight{1.25}\smash{\begin{tabular}[t]{l}O\end{tabular}}}}%
    \put(0.71653449,0.1380849){\color[rgb]{0,0,0}\makebox(0,0)[lt]{\lineheight{1.25}\smash{\begin{tabular}[t]{l}O$_2$\end{tabular}}}}%
  \end{picture}%
\endgroup%

    \input{FIGURES/neutrals_cb_1.pgf}
    \caption{Number density of neutral species (top: N$_2$, middle: O, bottom: O$_2$) in the Air-Breathing system at steady-state.}
    \label{fig:neutralsdensity}
\end{figure}

The simulation for the neutral flow is run for a total simulation time of 10~ms, necessary to reach a steady-state. The resulting neutrals number density at steady-state are plotted in Figure \ref{fig:neutralsdensity}. The number density of N$_2$ in the thruster ionization chamber reaches $10^{20}$~m$^{-3}$, the density of O$_2$ reaches $1.9\times 10^{19}$~m$^{-3}$, and that of O reduces to approximately $2\times 10^{16}$~m$^{-3}$, resulting in a compression ratio of 475 for nitrogen with respect to the free stream density, and of 90 for molecular oxygen with respect to the free stream density of atomic oxygen. The higher magnitude of the first value with respect to those reported in literature, typically in the range 100-300, can be explained by the favourable CLL gas-surface interaction kernel that promotes forward transmissivity of free stream particles through incomplete tangential accommodation while impeding the backwards transmissivity of particles that have already been scattered.

For simplicity purposes, in this work only nitrogen is considered as the operating gas in the simulation of the thruster. Therefore, a uniform background density of 10$^{20}$~m$^{-3}$ for N$_2$ is used in the subsequent simulations.

\subsection{Ionization chamber} \label{sec:ionizchamber}

\begin{figure*}[ht]
    \centering
    \input{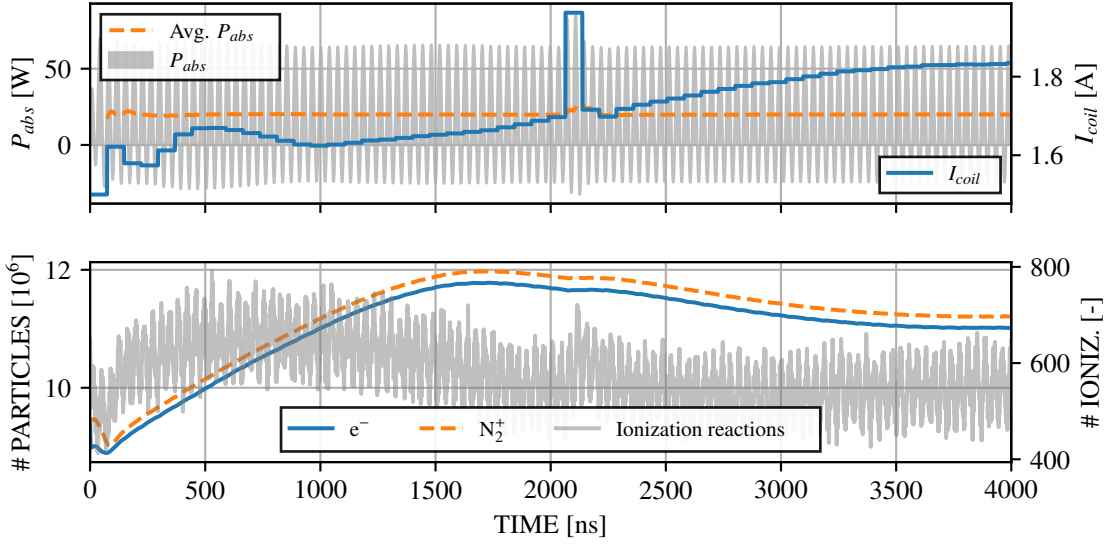}
    \caption{Relevant quantities in the ionization chamber. The top plot shows the power absorbed by the plasma from the RF electric field. The value averaged over one RF period is shown with the dashed line, and is found to be very close to the target set point of 20~W. The coil current $I_\text{coil}$ is also shown. This quantity is adjusted approximately every RF cycle, or 737 time steps according to Eq.~(\ref{eq:icoil}). The lower plot shows the total number of electrons and ions in the simulation, as well as the count of ionization events per time step. This quantity oscillates due to the periodic motion of the electrons.}
    \label{fig:chamber_quantities}
\end{figure*}

The simulation of the plasma flow in the thruster is run for a total physical time of 4~\textmu s, most of which were necessary to reach a quasi-steady state. Some quantities of interest in the ionization chamber, the power absorbed by the plasma $P_\text{abs}$, the current fed to the RF coil $I_\text{coil}$ are plotted in Figure~\ref{fig:chamber_quantities}. We've found the simple strategy to control the current applied to the coil, according to Eq.~(\ref{eq:icoil}), to be very effective in maintaining the absorbed power $P_\text{abs}$ close to the target value of 20~W. The absorbed power, averaged using a moving average filter of width equal to the RF period, is shown with the dashed line in the top plot of Figure~\ref{fig:chamber_quantities}. Except for a brief disturbance at around 2~\textmu s, due to a mistake in the configuration after a simulation restart, the coil current increases towards its steady-state value of 1.83~A.

\begin{figure}[ht]
    \centering
    \input{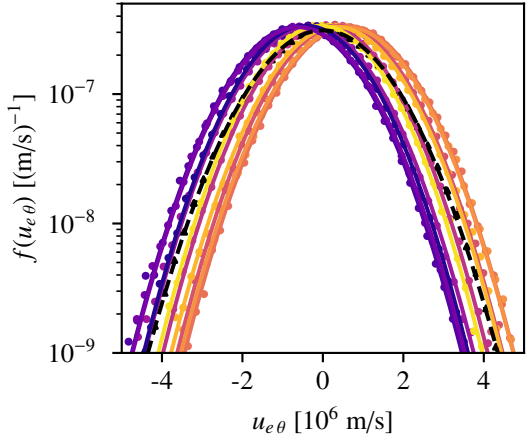}
    \caption{Circular markers: VDF for azimuthal electron velocity at different phases of the RF cycle. Lines: corresponding fits of Maxwell-Boltzmann distribution functions. Triangular markers: time averaged VDF. Dashed line: corresponding M-B fit.}
    \label{fig:eedf_mb}
\end{figure}

\begin{figure}[ht]
    \centering
    \input{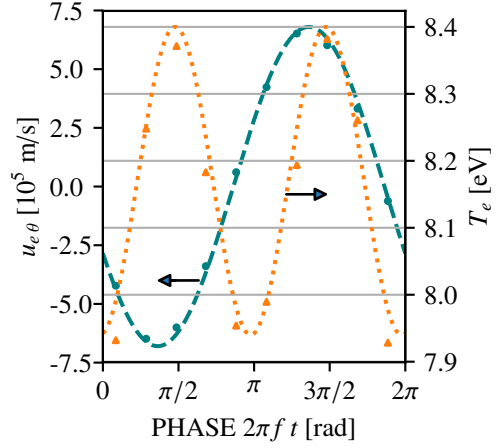}
    \caption{Markers: sampled azimuthal drift velocity and azimuthal traslational temperature at ten different phases of the RF cycle. Lines: approximate fits of sinusoidal functions to the samples.}
    \label{fig:eedf_uandt}
\end{figure}

The azimuthal velocity distribution of electrons in the external layer ($15<r<20$~mm) of the ionization region is shown by the circular markers in Figure~\ref{fig:eedf_mb}, from particles sampled at ten different phases of the RF cycle. The continuous lines are fits of the Maxwell-Boltzmann velocity distribution function. We notice that the sampled distribution is indistinguishable from a M-B distribution, drifting alternatively in the azimuthal direction. The black triangular markers and dashed line represent a time-averaged distribution, which has zero drift velocity but a higher effective temperature, measured at 9.4~eV. This distribution is also very closely Maxwellian. This is probably due to the relative magnitude of the elastic to ionization cross section at the relevant energies, see Figure~\ref{fig:crossections}. For clarity, the drift velocity and temperature obtained from the M-B fits are shown in Figure~\ref{fig:eedf_uandt}. Sinusoidal functions with frequency $f$ and $2f$ are approximately fit to the drift velocity and temperature samples, respectively. These are found to be $u_{e \, \theta} = -6.8\times 10^5 \, \cos(\omega t-1.15)$~m/s and $T_e = 8.17+0.23 \, \cos(2 (\omega t - 1.5))$~eV, with $\omega = 2\pi f$. The velocity therefore follows the electric field oscillation with a phase lag due to electron inertia, while the temperature increases following the electric field magnitude with a phase lag of approximately 90 deg. The alternating drift motion results in an effective electron temperature higher by about 1~eV with respect to the highest instantaneous temperature.

\subsection{Thruster plume} \label{sec:thrusterplume}

\begin{figure*}[ht]
    \centering
    \input{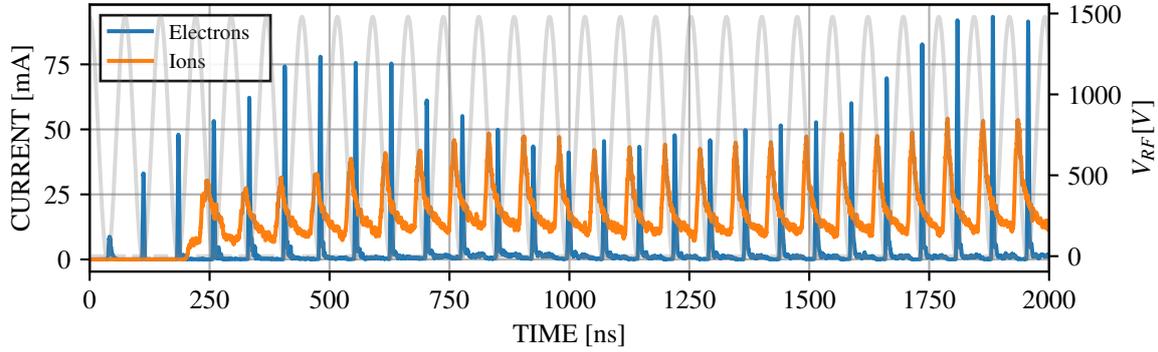}
    \caption{Electron and ion currents collected at the outlet surface of the thruster. A moving average filter over 20 time steps is applied to the instantaneous data to remove the noise. The screen grid potential $V_\text{RF}$ is shown in gray. The electron current pulse corresponds to the low potential phase of the cycle. Ions show a constant flow rate with periodic peaks. Propagation of the fastest ions from the acceleration grid to the outlet plane where current is measured takes approximately 170~ns, which causes the initial delay in the establishment of the ion current.}
    \label{fig:boundary_currents}
\end{figure*}

\begin{figure*}[ht]
    \centering
    \input{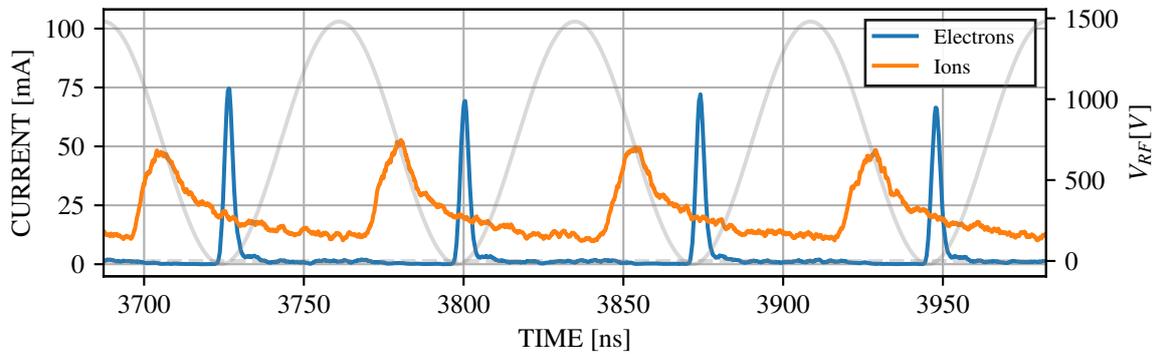}
    \caption{The same plot as that of Figure~\ref{fig:boundary_currents}, but for only four RF cycles in quasi-steady state conditions.}
    \label{fig:boundary_currents_ss}
\end{figure*}

\begin{figure}[ht]
    \centering
    \input{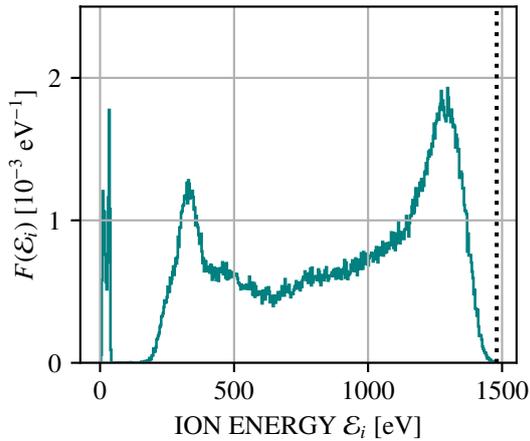}
    \caption{Ion energy flux distribution function measured at the outlet surface of the thruster. The vertical line shows the maximum potential between the grids during the RF cycle (1450~V).}
    \label{fig:iefdf}
\end{figure}

Figure~\ref{fig:iefdf} shows the axial energy flux distribution function of ions that reach the thruster outlet plane. The distribution presents two peaks, at approximately 350~eV and 1250~eV. As the authors illustrate in \cite{lafleur2019}, this distribution depends on the relative magnitude of the ion transit time through the grids and the RF period. The situation encountered here resembles the low- to intermediate frequency regime, which occurs at around 8~MHz in the specific conditions of \cite{lafleur2019}. This operational point is of course modified here because of the different ion mass, grid potential, and grid spacing. The shape of the distribution can be explained considering the fact that ion transit time through the grid is slower than the RF period, therefore ions enter the acceleration region during different phases of the RF cycle, experiencing different acceleration potentials. The energy is limited by the maximum potential difference present on the grids, and is also limited on the low side because the ions take a few nanoseconds to traverse the sheath on the screen grid once the potential barrier allows them to. The low energy peak in the distribution is due to ionization of the background neutrals in the screen and plume region by electrons. Notice that here we do not consider charge exchange collisions, which could also be a source of these low energy ions. The average thrust measured by integrating the momentum flux of the ions is 480~\textmu N.

\begin{figure}[ht]
    \centering
    \input{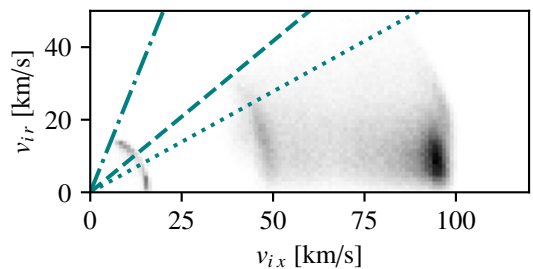}
    \caption{Axial-radial ion flux distribution function measured at the outlet surface of the thruster. The lines show the approximate divergence angles of 30, 40, and 70 deg from the axial direction for the three prevalent ion populations.}
    \label{fig:2difdf}
\end{figure}

Figure~\ref{fig:2difdf} shows the axial-radial velocity flux distribution function of the ions reaching the thruster outlet plane. The same populations of ions of Figure~\ref{fig:iefdf} are visible here as well. The three distinct peaks in the distribution produce plumes with divergence of approximately 30, 40, and 70 deg respectively.

\subsection{Comparison to the global model} \label{sec:globalmodel}

The global thruster model of \cite{chabert2012} can be further simplified by fixing the neutral gas density and temperature. The equations then reduce to:
\begin{equation}
\begin{cases}
    n n_g k_{iz} = n u_B \frac{A_\text{eff}}{V}\\
    \frac{\mathcal{P}_\text{abs}}{V} = E_{iz}n n_g k_{iz} + 3 \frac{m_e}{m_i} k_B (T_e-T_g)n n_g k_{el} \\ \quad + 7 k_B T_e n u_B \frac{A_\text{eff}}{V}
    \end{cases}
\end{equation}
Where $n=n_e=n_i$ is the average plasma density, $n_g$ and $T_g$ are the density and temperature of the background neutral gas, $u_B=(k_B T_e/m_i)$ is the Bohm velocity, $A_\text{eff}$ and $V$ are the surface area and the volume of the ionization chamber, $k_{iz}$ and $k_{el} = 10^{-13}$~m$^{-3}$ are the reaction rates for ionization and elastic electron-neutral interactions, and $E_{iz}=15.581$~eV is the ionization energy. The first equation yields the electron temperature, which only depends on the ionization cross section. The second equation can be solved for the plasma density. Imposing $\mathcal{P}_\text{abs}=20$~W, $n_g=10^{20}$~m$^{-3}$, $T_g=300$~K, and the following Arrhenius fit to the ionization cross section:
\begin{equation}
    k_{iz} = 1.322\times 10^{-18} \, T_e^{0.895} \exp(-179874.3 \text{K}/T_e) \text{ m$^3$ s$^{-1}$}
\end{equation}
Results in $T_e=7.6$~eV and $n=1.57\times 10^{17}$~m$^{-3}$. The peak plasma density in the PIC simulation is $n=1.65\times 10^{17}$~m$^{-3}$, and the averaged electron temperature is $T_e=7.8$~eV. Notice that the latter value is lower than what previously reported in Section~\ref{sec:ionizchamber} because it considers all velocity directions (not only the azimuthal one) and locations in the ionization chamber. We conclude that the simple global model represents well the energy and particle loss mechanisms.

\section{Conclusion and perspectives}

We have demonstrated a one-way coupled method for a fully kinetic simulation of an Air-Breathing Electric Propulsion system. We designed the geometry of the system for this work in order to be representative and according to engineering judgment, but we made no attempt to optimize its performance. Despite this, assuming the chosen model for particle reflections at the walls, we measured a compression ratio of 480  for nitrogen and of 84 for oxygen with respect to the free stream, with the desirable characteristic of rather uniform density in the thruster ionization chamber. We then fixed a neutral background density for nitrogen of $10^{20}$~m$^{-3}$ in the simulation of the thruster, since this density would allow to sustain a plasma in the target RF power range. The simple controller for the current applied to the RF coil was seen to be very effective in maintaining the power absorbed by the plasma close to the target of 20~W. The simulations of the thruster reached quasi-steady-state after approximately 3.5~\textmu s.

We analyzed the electron velocity distribution function in the azimuthal direction, in which the external electric field acts. We have observed it follows closely a Maxwell-Boltzmann velocity distribution function, with a periodic azimuthal drift due to the applied field. In correlation with the the phase of maximum drift velocity, the electron temperature also has a maximum, probably due to the effect of electron-neutral collisions converting part of the drift energy into random thermal energy. We compared the plasma density and electron temperature at steady-state with the result of a global mass and power balance in the ionization chamber, and found them to be in very good agreement. The measured electron and ion current showed the expected alternate emission of ions and electrons, with very good qualitative match to published 2D PIC simulation results where currents are measured at close distance from the acceleration grid. In experiments, the measurement is performed much further away from the grids, therefore the currents show different temporal profiles. Extension of the simulation to include a larger plume region could be considered in the future, in order to compare to the available experimental results. An analysis of the ion velocity flux distribution showed a main beam with large velocity spread, matching well to previously published results for this type of thrusters in the low- to intermediate frequency range. A population of slow ions is also present, probably generated by ionization in the acceleration and near plume regions.

We measured the thrust of the system by averaging the ion momentum flux at the outlet, to 480~\textmu N. The drag of the satellite, considering the free stream density of nitrogen as set in Section~\ref{sec:domain}, and a drag coefficient $C_D=2$, is of 5.7~mN. Optimization of the system, as well as a more efficient confinement of the plasma in the ionization chamber, could certainly improve the performance.

In future work, we plan to improve the modeling by including oxygen species, and include more complex collisional and reactive processes. In particular, the presence of negative oxygen ions is expected to play a significant role in the plasma chemistry and dynamics of the acceleration process.

\section*{Acknowledgments}

The research of PP is founded by an FWO Strategic Basic PhD fellowship (reference 1S24022N).

\bibliographystyle{vkiarticle-num}
\bibliography{vki}

%% Authors are advised to submit their bibtex database files. They are
%% requested to list a bibtex style file in the manuscript if they do
%% not want to use model5-names.bst.

%% References without bibTeX database:

% \begin{thebibliography}{00}

%% \bibitem must have one of the following forms:
%%   \bibitem[Jones et al.(1990)]{key}...
%%   \bibitem[Jones et al.(1990)Jones, Baker, and Williams]{key}...
%%   \bibitem[Jones et al., 1990]{key}...
%%   \bibitem[\protect\citeauthoryear{Jones, Baker, and Williams}{Jones
%%       et al.}{1990}]{key}...
%%   \bibitem[\protect\citeauthoryear{Jones et al.}{1990}]{key}...
%%   \bibitem[\protect\astroncite{Jones et al.}{1990}]{key}...
%%   \bibitem[\protect\citename{Jones et al., }1990]{key}...
%%   \harvarditem[Jones et al.]{Jones, Baker, and Williams}{1990}{key}...
%%

% \bibitem[ ()]{}

% \end{thebibliography}

\end{document}